# Title

Spectroscopies and electron microscopies unravel the origin of the first colour photographs

# Authors list

Victor de Seauve[a,b*], Marie-Angélique Languille[a*], Mathieu Kociak[c], Stéphanie Belin[d], James Ablett[d], Christine Andraud[a], Odile Stéphan[c], Jean-Pascal Rueff[d, e], Emiliano Fonda[d], Bertrand Lavédrine[a]

[a] Centre de Recherche sur la Conservation (CRC), Muséum national d'Histoire naturelle, CNRS, Ministère de la Culture, 36 rue Geoffroy Saint Hilaire, 75005 Paris, France.

[b] SACRe (EA 7410), Ecole normale supérieure, Université PSL, 75005 Paris, France.

[c] Laboratoire de Physique des Solides, Université Paris-Sud, CNRS UMR 8502, Orsay, France.

[d] Synchrotron SOLEIL, L'Orme des Merisiers - St. Aubin-BP 48, 91192 Gif s/Yvette, France.

[e] Sorbonne Université, CNRS, Laboratoire de Chimie Physique-Matière et Rayonnement, 75005 Paris, France

# Corresponding authors

* Marie-Angélique Languille, malanguille@mnhn.fr

* Victor de Seauve, victor.de-seauve@cicrp.fr

# Abstract

The first colours photographs were created by a process introduced by Edmond Becquerel in 1848. The nature of these photochromatic images colours motivated a debate between scientists during the XIX[th] century, which is still not settled. We present the results of chemical analysis (EDX, HAXPES and EXAFS) and morphology studies (SEM, STEM) aiming at explaining the optical properties of the photochromatic images (UV-visible spectroscopy and low loss EELS). We rule out the two hypotheses (pigment and interferences) that have prevailed since 1848, respectively based on variations in the oxidation degree of the compound forming the sensitized layer and periodically spaced photolytic silver planes. A study of the silver nanoparticles dispersions contained in the coloured layers showed specific localizations and size distributions of the nanoparticles for each colour. These results allow us to formulate a plasmonic hypothesis on the origin of the photochromatic images colours.

# Keywords

Edmond Becquerel, photochromism, Ag@AgCl, electron microscopy, EELS-LL

# License

# Introduction

Advances in physics of light during the XIX[th] century helped to develop some of the most interesting photographic processes such as Ducos du Hauron's three colour process[1], or Gabriel Lippmann's process of interferential photography.[2] Conversely, the use of silver halides in photography since the first half of the XIX[th] century led to intense research on their physico-chemical properties and their interaction with photolytic silver nanoparticles.[3,4] More recently, the plasmonic properties of silver nanoparticles at the surface of daguerreotypes have been studied comprehensively with modern analytical techniques.[5] Research on silver halides is experiencing a strong resurgence of interest due to applications in photochromic devices[6,7] and in plasmonic photocatalysis.[8,9] Basing himself on Thomas Johann Seebeck's[10] and John Herschel's[11] work on silver chloride, Edmond Becquerel developed in 1848 the first colour photographic process[12]. These first colour photographs motivated a long-lasting debate through the XIX[th] century between the supporters of a pigment hypothesis on the origin of these colours, the most famous being William de Wiveleslie Abney and Mathew Carey Lea and those who postulated an interferential origin to the colours of the photochromatic image, such as Wilhelm Zenker[13] and Otto Wiener[14]. Recent work allowed to reproduce Becquerel process in the laboratory[15,16] and to show that the sensitized layer, which is coloured during the exposition to light, is made of silver nanoparticles (NP) dispersed in a silver chloride grains matrix[15,17]. However, until now light has still not been shed on the nature of the photochromatic images colours.

Gabriel Lippmann did not believe that the photochromatic images colours were due to interferences[18]; Edmond Becquerel[19] and his son Henri Becquerel[20] agreed as well. After having adapted Becquerel's process on glass and having obtained colours that were similar in reflection and transmission, Abney stated that these were not interferential and that they were due to oxidation and reduction products of the silver salt constituting the sensitized layer.[21–23] For Carey Lea, along with Becquerel, the latter was constituted by "*silver photochloride*", a mixture of silver chloride and "*silver subchloride*" of chemical formula "$Ag_2Cl$".[12,24] Carey Lea also explained that the silver photochloride colour could change depending on the silver chloride/silver subchloride ratio.[24] Quite early, Zenker postulated that the photochromatic images colours were due to a series of parallel and periodically spaced photolytic silver planes, which would create standing waves and interferential colours.[13,25] This hypothesis was also put forward by John Rayleigh[26] and Otto Wiener. The latter reproduced Becquerel's process and noticed that the colours he obtained changed when viewed through a prism, which proved that these were interferential colours.[14] Only after Gustav Mie's publication on the absorption of light by nanoparticles[27] a third theory was put forward by Hinricüs Lüppo-Cramer: the *silver photochloride*-based sensitized layer would be reduced by the exposition to light into silver nanoparticles of one specific size, leading to one colour[28] (quoted in [29]). Lastly, Céline Dupont *et al.* measured by X-ray photoemission spectroscopy (XPS) modern *heliochromies* obtained by Abel Niépce de Saint-Victor's adaptation of Becquerel's process and noticed differences in the oxidation of copper between the colours; they concluded that the latter were due to copper complexes at the surface of the images.[30,31]

Few recent publications directly relate to Becquerel's photochromatic images. Nevertheless, we identified the sensitized layer as being constituted by silver nanoparticles and silver chloride[17], which makes it comparable to other photochromic systems. Ageev et al. measured a decrease in absorption around the visible exposition wavelength when irradiating silver chloride crystals containing silver nanoparticles dispersions. They attributed this phenomenon called *spectral photoadaptation* to the modification of the nanoparticles sizes and distances between each other.[7,32] The loss of absorption at the visible exposition wavelength is a phenomenon that has also been observed on photolyzed silver chloride crystals by Hilsch and Pohl[33], Brown and Wainfan[34], and Moser *et al.*[35] and explained by Hughes by the selective *dissolution* of one specific size of silver nanoparticles.[36] The latter hypothesis was also put forward by Matsubara and Tatsuma in order to explain the photochromism of a system based on silver nanoparticles dispersed on a $TiO_2$ surface.[6] A silver nanoparticle absorbs light by surface plasmon resonance (SPR) and the energy of this resonance depends on the nanoparticle size, shape and environment.[27,37,38] A loss of a certain type of silver nanoparticles in Matsubara and Tatsuma's system thus results in an absorption loss at a specific wavelength.

In this context, the XIX[th] hypotheses on the origin of the photochromatic images have to be revisited. We thus analysed several sensitized and coloured samples produced by our replication of Becquerel's process in the laboratory.[16] We will first present a study of the pigment hypothesis which involves chemical analysis of coloured layers produced by the Becquerel process, then we will discuss the interferential hypothesis in light of results on the morphology of these coloured layers. Finally, we will show that the silver nanoparticles dispersions contained in the coloured layers could be at the origin of the photochromatic images colours in the frame of a new plasmonic hypothesis. Note that a previous study of the X-ray and electron beam effects allowed us to determine controlled or beam damage free conditions for these analysis.[17]

# Results and discussion

<u>Pigment-based hypothesis surveyed by spectroscopies.</u> We previously showed that both the volume and the surface of sensitized layers are constituted by Ag, Cl and traces elements [17]. Energy dispersive X-ray spectroscopy (EDX) and Hard X-ray photoelectron spectroscopy (HAXPES) analysis performed on the GALAXIES beamline[39] of the SOLEIL synchrotron showed that the sensitized layer and all coloured samples have the same composition (see figures S-1 and S-2 in the Supporting Information (SI) section). In model samples, no copper has been detected, neither by means of EDX, HAXPES nor even by X-ray Absorption Spectroscopy (XAS) at the Cu K-edge. These results discard the Dupont *et al.* theory that involves copper complexes at the surface of photochromatic images that would be responsible for their colours.[30,31]

Figure 1 shows the valence band spectra of several coloured real samples. Since the slight shift of the valence band from red to blue was observed on all the measured levels and Auger peaks, we attributed it to a variation in the sensitized layers thicknesses; charge effects were indeed reported by Mason, who analysed AgCl layers as thin as 40 nm in order to avoid them.[40] The valence bands of all coloured samples are thus comparable. These results, along with the XAS results obtained on the ROCK beamline[41] of the SOLEIL synchrotron (see figures S-3 for the XAS spectra and S-4 for the FTs of the Extended X-ray absorption fine structure (EXAFS) signals in the SI section) show that the sensitized layer does not undergo chemical change during the colouration. This discards the *photochloride* pigment hypothesis put forward by Abney[21–23] and Carey Lea.[22]

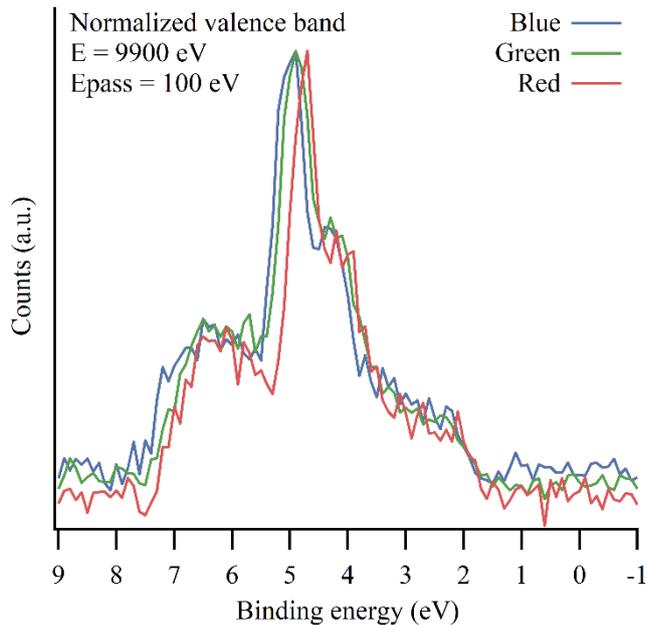

Figure 1. Valence band spectra measured by HAXPES in three coloured samples (sensitized silver deposits on silicon wafers). The zero of binding energy was determined from the Fermi level of pure silver.

<u>Interference-based hypothesis studied by electron microscopy (EM) and UV-visible spectroscopy.</u> We previously showed that the coloured layers chemical composition is the same as the sensitized layers. Studies in scanning electron microscopy (SEM) and scanning transmission electron microscopy STEM of the morphology of coloured layers, both in plane and cross-section view, showed that the latter are constituted, just as the sensitized layers[17], by silver nanoparticles of sizes ranging from a few nanometres to 150 nm dispersed in a micrometric silver chloride grains matrix. These nanoparticles were located at silver chloride grains edges, boundaries and in their volumes (for examples, see figures 8 and 9 in [17]). They did not form any periodic structures, neither at coloured layers surface, nor in their volume. Furthermore, figure 2 presents the UV-visible absorptance spectra of several coloured model samples; absorptance holes appear in these spectra at the exposition wavelengths, which indicate that the colours are the results of an absorption process. These results discard the hypothesis put forward by Zenker[13,25] and Wiener[14] of periodic parallel silver planes in the volume of the coloured layers which would results in standing waves and interferential colours. Note that, even if the surface morphologies at the [100 nm; 10 μm] scale (see figures S-5 in the SI section) and the absorptance spectra (figure 2) show that colours are not due to the diffusion of light, it can play a role in the overall colour of the sensitized and coloured layer, because of their porosity and surface roughness.

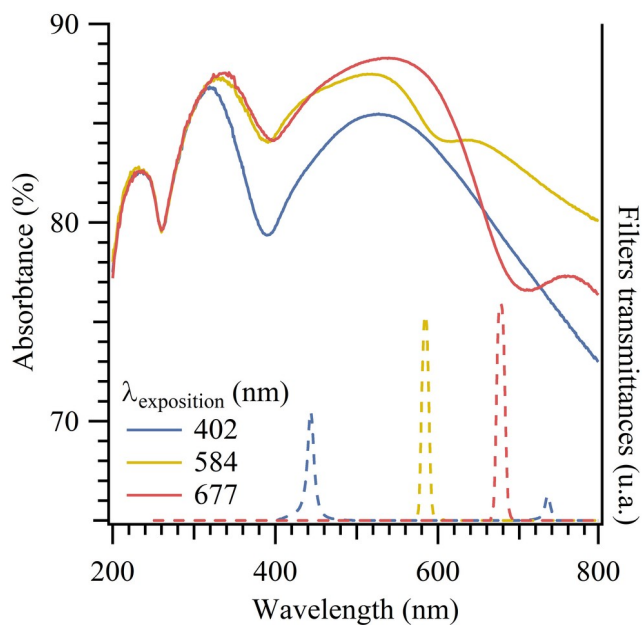

Figure 2. UV-visible absorptance spectra of coloured model samples.

<u>Toward a plasmonic hypothesis.</u> Using large sets of high magnification STEM images, the silver nanoparticles contained in coloured layers were classified, according to their size, localization in relation to silver chloride grains, orientation in relation to the layer's surface and shape. It appears that the orientation and the shape distributions of the silver nanoparticles are the same for all coloured samples. However, their localization and size distributions differ, as figure 3 shows. The histograms represent the silver nanoparticles size distribution variation for coloured samples against the size distribution of a sensitized sample; a positive (respectively a negative) value thus represents an increase (respectively a decrease) of the relative quantity of particles which sizes are within a given range. We see that for the blue sample, the overall number of particles has increased, especially for the [5; 30] nm range. For the yellow sample, the number of [5; 25] nm nanoparticles has increased but the number of [25; 45] nm nanoparticles has decreased. For the red sample, it seems that nanoparticles smaller than 25 nm have appeared and that [30; 45] nm nanoparticles have disappeared. The same kind of histograms for nanoparticles specifically located inside silver chloride grains or at grain edges are available in the SI section (figures S-6); it appears that mostly embedded nanoparticles disappeared for the yellow samples, whereas mostly grain edges nanoparticles variations disappeared for the red sample.

The overall increase of the number of nanoparticles in the blue sample is confirmed by figure S-7a in the SI section, which shows the total number of silver nanoparticles per unit of layer length for sensitized and coloured samples. It is interesting to compare this result to the increase of the relative volume occupied by the nanoparticles in the blue sample versus that in the sensitized sample (figure S-7b in the SI section). This can be interpreted as follow: during the exposition of the sensitized layer to blue light of peak wavelength 446 nm, silver chloride contained in the sensitized layer is photolyzed and silver appears under the form of new nanoparticles. Indeed, even if AgCl absorption edge is located around 404 nm at ambient temperature[17,42], it can be self-sensitized by the presence of silver nanoparticles[43,44]; photolysis was observed at 473 nm exposition wavelength but not at 502 nm. For yellow and red samples, for which the total amount of silver remains constant (figure S-7b in the SI section), the increase or decrease of the number of nanoparticles in some size ranges is then due to the reorganization of the nanoparticles size distribution: some nanoparticles get smaller, in favour of others that get bigger.

According to the size distributions (figure 3), bigger nanoparticles are expected when going from blue to red, which is confirmed when looking at figure S-7c in the SI section showing the average size of silver nanoparticles in several coloured samples. This is in apparent contradiction with a plasmonic origin of these colours since bigger silver nanoparticles are expected to absorb at lower energies[27]. Nonetheless, in order to understand the colours, the localization and size distributions are more important: we suggest that the samples colours are due to a loss of silver nanoparticles of some specific size range (e.g. [30; 45] nm for the red sample, figure 3c) and localization (mostly grain edges nanoparticles, figure S-4e), resulting in an absorption hole located around the exposition wavelength (around 676 nm for the red sample, figure 2). In the same way the increase of absorption in the visible region, except around the exposition wavelength, would be due to the increase of the number of nanoparticles out of these localization and size ranges.

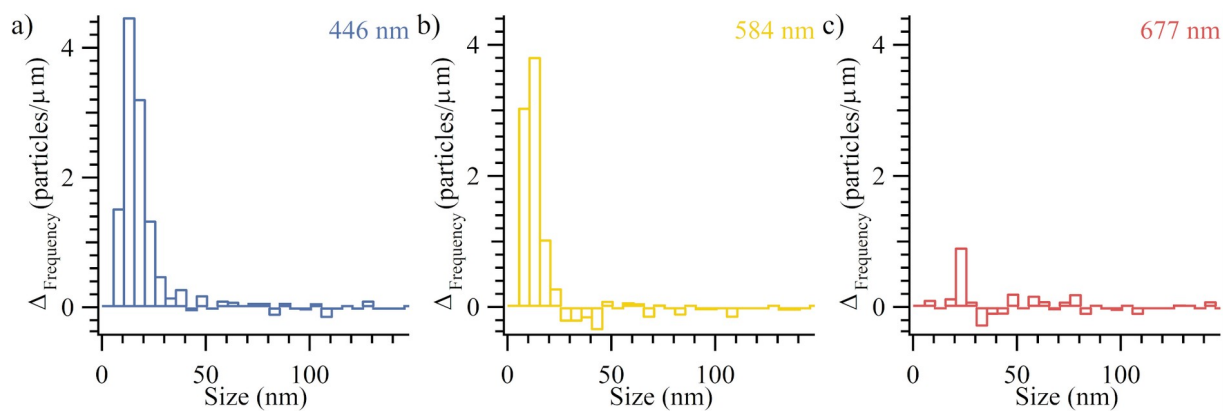

Figure 3. Ag-NPs size distribution variations for several coloured samples obtained by subtracting the frequencies of the Ag-NPs size distribution of a sensitized sample to that of coloured samples.

SPR around silver nanoparticles studied by low-loss electron energy loss spectroscopy (EELS). In order to study the SPR modes of the silver nanoparticles contained in coloured samples, low-loss EELS maps of the latter, under the form of Spectrum-images (SPIM), were acquired. Figure 4b shows the high angle annular dark field (HAADF) image of a silver nanoparticle located on a silver chloride grain edge contained in a blue sample; figure 4a shows two EELS spectra extracted from the corresponding circled area on figure 4b. Two resonance modes are identified on these spectra: around 1.9 eV (light green) and 2.8 eV (yellow). Figures 4c and 4d show the EELS intensity maps filtered around these energies; the two modes are clearly located at the interface between the silver chloride grain and the silver nanoparticle for the lower energy mode (figure 4c) and around the side of the nanoparticle facing the void for the higher energy mode (figure 4d). The SPR energy of a nanoparticle shifts to higher energy when the refractive index of the surrounding medium decreases[37]; two SPR modes were thus expected for a nanoparticle, on side of which being in contact with silver chloride and the other with void. Wei *et al.* already measured several SPR modes in the vicinity of silver nanoparticles deposited on ZnO nanowires[45].

The SPR modes of a dozen of grain edge nanoparticles in three coloured samples have been measured. Their energies have been plotted against the nanoparticles sizes (figure 4e). Void side modes range between 2.7 and 3.2 eV whereas silver chloride grain side modes range between 2.3 and 2.8 eV. For both modes, the energy decreases as the nanoparticles size increases, which is in agreement with the predictions; indeed, as the plasmon is confined at the nanoparticles surface, when the size of the latter increases, the plasmon wavelength does the same, and the SPR peak shifts toward lower energies[27,38]. Grain edge nanoparticles absorb incident light between 2.3 and 3.2 eV; besides, some nanoparticles embedded in silver chloride grains were studied and SPR modes were measured down to 1.8 eV. These energies correspond to the visible region and these results show that the silver nanoparticles contained in the sensitized and coloured layers can be at the origin of the photochromatic images colours.

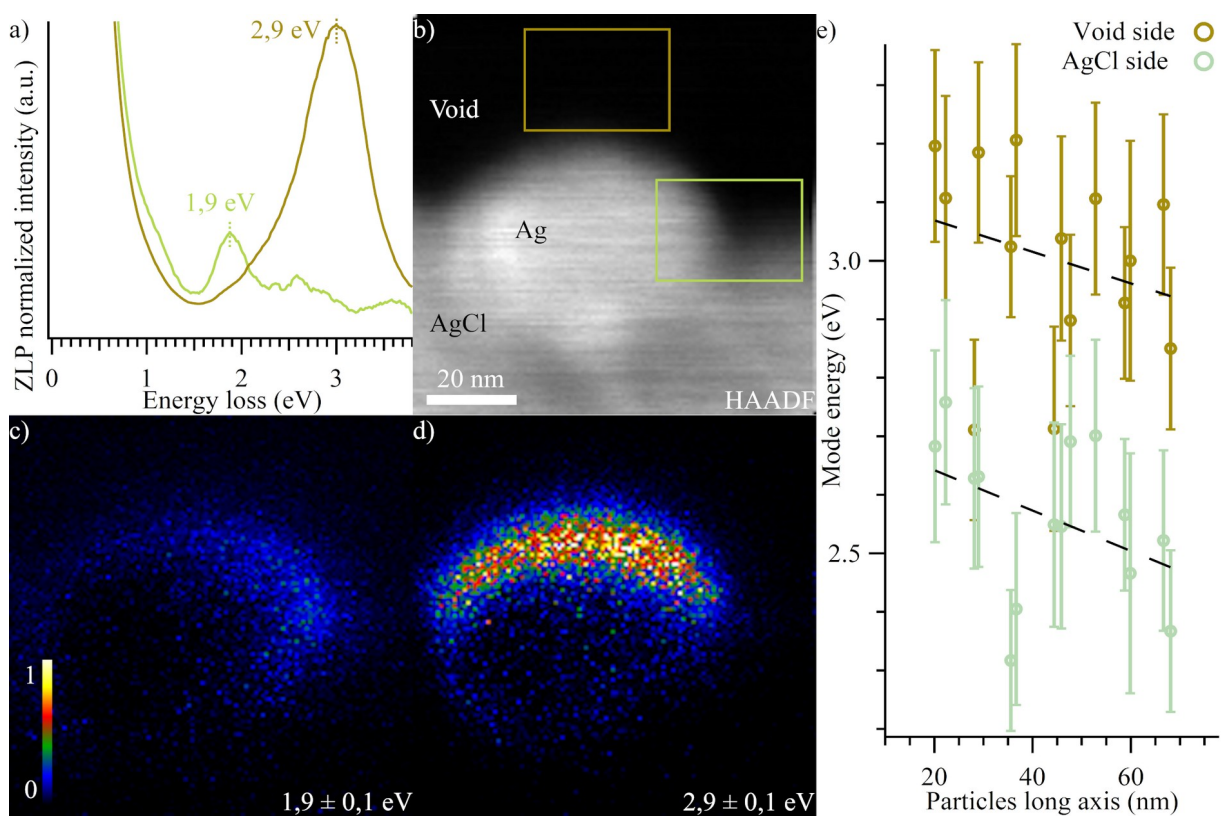

Figure 4. a) Zero Loss Peak (ZLP) normalized energy loss spectra extracted from the areas represented on the HAADF image (b) of a silver nanoparticle located on an AgCl grain edge. c) and d) Normalized EELS intensity maps obtained from the spectrum image for two resonance modes. e) Energy positions of the "AgCl side" and "void side" modes measured around several nanoparticles versus their long axis size.

Photoadaptation mechanism during the exposition step. We postulate that during the exposition of sensitized layers, two mechanisms coexist. Firstly, when exposing layers with sufficiently energetic light, silver chloride is partially photolyzed in a dispersion of silver nanoparticles. Secondly, silver nanoparticles of specific localizations and size ranges which absorb the incident radiation by SPR are modified by the exposition; their size can be reduced, shifting their SPR peak toward higher energy, and silver can be transferred to other nanoparticles, thus increasing their size. This results in the decrease of the number of particles absorbing the incident radiation and thus in a coloured layer displaying an absorption hole around the exposition wavelength. The appearance of an absorption hole is less clear for lower exposition wavelength (see the blue spectrum on figure 2), probably because the photolysis that also takes place during the exposition reshapes the global diffusion/absorption ratio of the layers.

We suggest for the photoadaptation a process close to what Matsubara and Tatsuma described in [6], in order to explain the photochromism of an Ag-NP/TiO$_2$ system. The excitation of plasmons by visible photons leads to the promotion of conduction band electrons of the silver nanoparticles either to the conduction band of the semiconductor (AgCl in our case), either to the oxygen surrounding the nanoparticles; the great mobility of Ag$^+$ ions in a silver chloride matrix[3,4] allows them to reach available nanoparticles, or to form new particles by coalescence and recombination with an electron. Note that the energetically favoured transfer of electrons to oxygen surrounding the nanoparticles could be a reason for the selected loss of grain edge nanoparticles (figure S-4d e and f). The role of the oxygen as a charge carrier is still to be elucidated, as Ohko et al. showed no photochromic activity of a Ag/TiO2 system in an oxygen-free environment.[46] Computational techniques, which allowed Kelly and Yamashita to study charge transfers in Ag/TiO$_2$ systems with success[47], can be a great tool for this investigation, as first results already showed.[48] Note also that in the case of silver ions forming new nanoparticles by coalescence, crystalline defects are privileged locations[3,4], and that silver nanoparticles located at silver chloride grain boundaries in coloured layers were too small to be taken into account in this study; still, they can play a role in photochromatic images colours.

The comparison of photochromatic images with Ag/TiO$_2$ systems[46,49–52] is also supported by Kawahara et al.[53], who state that photochromism phenomena can virtually be observed with any semiconductor. As a first approximation, the Ag/AgCl Schottky barrier can be estimated at 2.8 eV[54] (versus 0.4 eV for TiO$_2$[50]), but metal induced gap states can significantly lower this barrier[55], allowing the creation of electron/holes pairs by visible photons. Besides, the mobility of silver ions in silver chloride allow them to directly cross the surrounding matrix; to the contrary, Matsubara et al. showed that a thin water layer at TiO$_2$ surface is necessary, in order for silver ion to move and to reach existing nanoparticles[56]. It is still important to note that the presence in silver chloride of silver atoms leads to a greater adsorption of silver clusters on AgCl than on TiO$_2$, as shown by Ma et al., and thus to specific interactions[57].

# Conclusion

Several coloured samples of photochromatic images were analysed in order to assess XIX$^{th}$ hypotheses on the origin of their colours. EDX, HAXPES and EXAFS showed no differences in chemical composition between various coloured samples. EM observations in plane view (SEM) and cross section view (STEM) did not show any differences in the morphology of coloured samples as well. UV-visible analysis of coloured samples displayed absorption holes around the exposition wavelengths used to colour the samples. The STEM study of silver nanoparticles dispersions contained in coloured samples showed that they are characteristic of the colours in terms of localizations and size distributions. Finally, low loss EELS maps showed that those nanoparticles absorb in the visible range, between 1.8 and 3.2 eV and thus could be responsible for the samples colours.

These results lead us to discard several hypotheses on the origin of the photochromatic images colours. The pigment hypothesis of Abney[23] and Carey Lea[24], and Dupont et al.[31] hypothesis were rejected in regard to the invariant chemistry and the absence of copper in coloured samples. The absence of periodic pattern, and the evidence that an absorptive process creates the colours in photochromatic images allowed us to discard the hypothesis of interferential colours put forward by Zenker[13] and Wiener.[14] Lastly, the evidence that the silver nanoparticles dispersions are modified by the exposition of sensitized layers to light made us suggest a plasmonic hypothesis of the photochromatic images colours. This hypothesis can also explain the phenomenon observed by Hilsh and Pohl[33], Brown and Wainfan[34] and Moser *et al.*[35] on photolyzed silver chloride crystals. A charge transfer mechanism at the interface between silver nanoparticles and silver chloride would lead to the transfer of silver from the nanoparticles which absorb the incident radiation by SPR to existing nanoparticles or new ones by coalescence. The resulting colour would be due to the subsequent lack of the type of nanoparticles which absorbed the incident radiation, causing the absorption hole around the exposition wavelength.

# Acknowledgement

The authors would like to thank the SACRe doctoral program (PSL University) for the financial support. This work was partially supported by the Observatoire des Patrimoines de Sorbonne Universités (OPUS). Thanks to the beamtime allocation n° 20150273, HAXPES experiments were performed on the GALAXIES beamline at SOLEIL Synchrotron, France. The authors wish to acknowledge the award of beamtime n° 20160485 on the ROCK beamline at Synchrotron SOLEIL which was supported by a public grant overseen by the French National Research Agency (ANR) as a part of the "Investissements d'Avenir" program ref: ANR-10-EQPX-45. The authors acknowledge financial support from the CNRS-CEA "METSA" French network (FR CNRS 3507) on the platform LPS-STEM.

# References

[1] M. Cotte, T. Fabris, J. Langlois, L. Bellot-Gurlet, F. Ploye, N. Coural, C. Boust, J.-P. Gandolfo, T. Galifot, J. Susini, *Angew. Chem. Int. Ed.* **2018**, *57*, 7364–7368.
[2] G. Lippmann, *Comptes Rendus Hebd. Séances Académie Sci.* **1891**, *112*, 274–275.
[3] J. Hamilton, *Adv. Phys.* **1988**, *37*, 359–441.
[4] T. Tani, in *Photogr. Sci. Nanoparticles J-Aggreg. Dye Sensitization Org. Devices*, Oxford University Press, **2011**, pp. 46–81.
[5] A. E. Schlather, P. Gieri, M. Robinson, S. A. Centeno, A. Manjavacas, *Proc. Natl. Acad. Sci. U. S. A.* **2019**, *116*, 13791–13798.
[6] K. Matsubara, T. Tatsuma, *Adv. Mater.* **2007**, *19*, 2802–2806.
[7] L. A. Ageev, V. K. Miloslavskii, Kh. I. El-Ashkhab, *Opt. Spectrosc.* **1992**, *73*, 213–216.
[8] P. Wang, B. Huang, X. Qin, X. Zhang, Y. Dai, J. Wei, M.-H. Whangbo, *Angew. Chem.-Int. Ed.* **2008**, *47*, 7931–7933.
[9] X. Zhang, Y. L. Chen, R.-S. Liu, D. P. Tsai, *Rep. Prog. Phys.* **2013**, *76*, 046401.
[10] T. J. Seebeck, in *Zur Farbenlehre*, J.G. Cottaschen, Tübingen, **1810**, pp. 716–720.
[11] John F. W. Herschel, *Philos. Trans. R. Soc. Lond.* **1840**, *130*, 1–59.
[12] E. Becquerel, *Ann. Chim. Phys.* **1848**, *22*, 451–459.
[13] W. Zenker, *Lehrbuch Der Photochromie (Photographie in Natürlichen Farben)*, Selbstverl., Berlin, **1868**.
[14] O. Wiener, *Ann. Phys.* **1895**, *291*, 225–281.
[15] V. de Seauve, A l'origine des couleurs des images photochromatiques d'Edmond Becquerel : étude par spectroscopies et microscopies électroniques, PhD Thesis, PSL Research University, **2018**.
[16] V. de Seauve, M.-A. Languille, S. Vanpeene, C. Andraud, C. Garnier, B. Lavédrine, *J. Cult. Herit.* **Submitted the 16/012020**, https://arxiv.org/abs/2001.05250.

[17] V. de Seauve, M.-A. Languille, S. Belin, J. Ablett, J.-P. Rueff, C. Andraud, N. Menguy, B. Lavédrine, *Anal. Chem.* **Submitted the 21/01/2020**, http://arxiv.org/abs/2001.07441.
[18] G. Lippmann, in *Prix Nobel*, Imprimerie Royale. P.A. Norstedt & Söner, Stockholm, **1909**.
[19] E. Becquerel, *Comptes Rendus Hebd. Séances Académie Sci.* **1891**, *112*, 275–277.
[20] H. Becquerel, *Les Expériences de M. Edmond Becquerel sur les actions chimiques de la lumière et l'héliochromie: Conférence du 20 mars 1892*, Gauthier-Villars, Paris, France, **1892**.
[21] W. de W. Abney, *Proc. R. Soc. Lond.* **1878**, *27*, 291–292.
[22] W. de W. Abney, *Proc. R. Soc. Lond.* **1879**, *29*, 190–190.
[23] W. de W. Abney, *Nature* **1896**, *54*, 125.
[24] M. Carey Lea, *Am. J. Sci.* **1887**, *33*, 349–364.
[25] W. Zenker, *Jahrb. Für Photogr. Reprod.* **1891**, *5*, 294–303.
[26] J. W. S. Rayleigh, *Philos. Mag. Ser. 5* **1887**, *24*, 145–159.
[27] G. Mie, *Ann. Phys.* **1908**, *330*, 377–445.
[28] H. Lüppo-Cramer, in *Ausführliches Handb. Photogr.*, W. Knapp, Halle-sur-Saale, **1927**.
[29] E. Mutter, in *Farbphotographie*, Springer Vienna, Vienna, **1967**, pp. 30–121.
[30] C. Dupont, M. Giraudet, S. Bourgeois, M. Kereun, J.-C. Niépce, B. Domenichini, Montpellier, **2014**.
[31] C. Dupont, M. Giraudet, S. Bourgeois, J. Fournier, M. Kereun, J.-C. Niépce, B. Domenichini, Paris, **2015**.
[32] L. A. Ageev, V. K. Miloslavskii, M. V. Verminskii, *Opt. Spectrosc.* **1997**, *83*, 148–153.
[33] R. Hilsch, R. W. Pohl, *Z. Für Phys.* **1930**, *64*, 606–622.
[34] F. C. Brown, N. Wainfan, *Phys. Rev.* **1957**, *105*, 93–99.
[35] F. Moser, N. R. Nail, F. Urbach, *J. Phys. Chem. Solids* **1959**, *9*, 217–234.
[36] A. E. Hughes, S. C. Jain, *Adv. Phys.* **1979**, *28*, 717–828.
[37] K. M. Mayer, J. H. Hafner, *Chem. Rev.* **2011**, *111*, 3828–3857.
[38] J. A. Scholl, A. L. Koh, J. A. Dionne, *Nature* **2012**, *483*, 421–427.
[39] J.-P. Rueff, J. M. Ablett, D. Céolin, D. Prieur, Th. Moreno, V. Balédent, B. Lassalle-Kaiser, J. E. Rault, M. Simon, A. Shukla, *J. Synchrotron Radiat.* **2015**, *22*, 175–179.
[40] M. G. Mason, *Phys. Rev. B* **1975**, *11*, 5094–5102.
[41] V. Briois, C. La Fontaine, S. Belin, L. Barthe, T. Moreno, V. Pinty, A. Carcy, R. Girardot, E. Fonda, *J. Phys. Conf. Ser.* **2016**, *712*, 012149.
[42] F. C. Brown, *J. Phys. Chem.* **1962**, *66*, 2368–2376.
[43] G. Calzaferri, D. Brühwiler, S. Glaus, D. Schürch, A. Currao, in *Proc. Intern Symp Silver Halide Technol.*, Hotel Mont Gabriel, Quebec, **2000**, pp. 59–62.
[44] S. E. Sheppard, *J. Frankl. Inst.* **1930**, *210*, 587–605.
[45] J. Wei, N. Jiang, J. Xu, X. Bai, J. Liu, *Nano Lett.* **2015**, *15*, 5926–5931.
[46] Y. Ohko, T. Tatsuma, T. Fujii, K. Naoi, C. Niwa, Y. Kubota, A. Fujishima, *Nat. Mater.* **2003**, *2*, 29–31.
[47] K. L. Kelly, K. Yamashita, *J. Phys. Chem. B* **2006**, *110*, 7743–7749.
[48] A. Lorin, M. Gatti, L. Reining, F. Sottile, Porquerolles, France, **2018**.
[49] K. Naoi, Y. Ohko, T. Tatsuma, *J. Am. Chem. Soc.* **2004**, *126*, 3664–3668.
[50] J. Okumu, C. Dahmen, A. N. Sprafke, M. Luysberg, G. von Plessen, M. Wuttig, *J. Appl. Phys.* **2005**, *97*, 094305.
[51] C. Dahmen, A. N. Sprafke, H. Dieker, M. Wuttig, G. von Plessen, *Appl. Phys. Lett.* **2006**, *88*, 011923.
[52] S. Glaus, G. Calzaferri, *J. Phys. Chem. B* **1999**, *103*, 5622–5630.

[53] K. Kawahara, K. Suzuki, Y. Ohko, T. Tatsuma, *Phys. Chem. Chem. Phys.* **2005**, *7*, 3851.
[54] A. W. Dweydari, C. H. B. Mee, *Phys. Status Solidi A* **1975**, *27*, 223–230.
[55] S. Glaus, G. Calzaferri, R. Hoffmann, *Chem. - Eur. J.* **2002**, *8*, 1785.
[56] K. Matsubara, K. L. Kelly, N. Sakai, T. Tatsuma, *Phys. Chem. Chem. Phys.* **2008**, *10*, 2263.
[57] X. Ma, Y. Dai, M. Guo, Y. Zhu, B. Huang, *Phys. Chem. Chem. Phys.* **2013**, *15*, 8722.

# Experimental section

<u>Real and model samples.</u> Figure S-8 in the SI section represents real and model samples. Edmond Becquerel's process of sensitizing silver layers has been replicated in the laboratory. The sensitization of *real* samples has been described in detailed in [16] and *model* samples have been introduced in [17]. Real samples are constituted by a sensitized 30 μm-thick silver deposit onto a 800 μm-thick brass support, whereas model samples consist in an integrally sensitized 1 μm-thick silver layer. The sensitized layers are a few microns thick pinkish photon- and electron-sensitive layers which have the ability to be coloured by incident visible light. Coloured samples are prepared by exposing sensitized layers to visible light, using narrow-bandwidth interferential filters, at a common radiant exposure of 10 kJ m$^{-2}$.[16]

<u>Synchrotron radiation based techniques.</u> The experimental conditions in which HAXPES and EXAFS data have been acquired on the GALAXIES[39] and ROCK[41] beamlines of the SOLEIL synchrotron have been described elsewhere.[17] Setups have been optimized in order to reduce the photon flux density onto the sample: by orienting it, continuously moving it under the beam and using filters in HAXPES and by working during the low intensity mode of the synchrotron machine in XAS. Moreover, the sample was cooled down to 200 K in HAXPES and the Quick-EXAFS monochromator of the ROCK beamline was used for fast data collection. In HAXPES, limited and controlled beam damages have been observed. In XAS, the samples were exposed to doses ranging from a few $10^{-4}$ GGy to a few $10^{-3}$ GGy and no beam damages have been observed.

<u>Electron microscopies.</u> The SEM and STEM devices that were used, the experimental conditions and the samples preparations have been described in [17]. The Zeiss in-column "In-Lens" secondary electrons (SE) detector is used in the SEM because of its excellent resolution at low voltage. The HAADF detector is used in the STEM for the chemical contrast it provides. SEM samples are real or model samples onto which a few micron thick carbon deposit has been made in order to protect them from the electron beam until doses as high as a few GGy. For the STEM, 70 nm thick sections of model samples have been cut by ultramicrotomy in a dark room in which the ambient light was regularly controlled. These sections are collected on carbon-coated Cu grids and sandwiched with a 5 nm thick carbon deposit in order to be protected from the electron beam. At liquid nitrogen temperature, no beam damages have been observed on C-sandwiched sections until doses as high as a few $10^3$ Ggy. Each sample has been imaged at low magnification and high magnification images of silver nanoparticles contained in the samples have been acquired, in order to get their size distributions. For each sample, 30 to 40 μm of layer has been studied, resulting in 300 to 600 nanoparticles measurements. Their major and minor axis were measured; their aspect ratios were then calculated and their sizes were calculated as the geometric means of their axis.

<u>UV-visible spectroscopy.</u> The UV-Visible spectroscopic measurements were performed on a CARY5000 spectrophotometer equipped with an integrating sphere in the total reflectance and transmittance modes (1 nm step, 1 s nm$^{-1}$). The absorptance was calculated as follows.

$$Absorbtance\,(\%) = 100 - Reflectance\,(\%) - Transmittance\,(\%)$$

<u>Low loss EELS.</u> Analysis have been made on the VG HB501 microscope of the Laboratoire de Physique des Solides (LPS, UMR8502, Orsay, France), working at 100 keV, 50 pA beam current and with a 50 μm condenser aperture. The convergence semi-angle is 7 mrad and the collection semi-angle is 7 mrad. The dispersion is 0.017 eV/channel, and the zero loss peak FWHM varied in the range of 0.241 to 0.369 eV. Spectrum images (SPIM) were aligned, background-subtracted and deconvoluted with a Richardson-Lucy algorithm using the Digital Micrograph® software and packages developed at the LPS.

# Supplementary Information

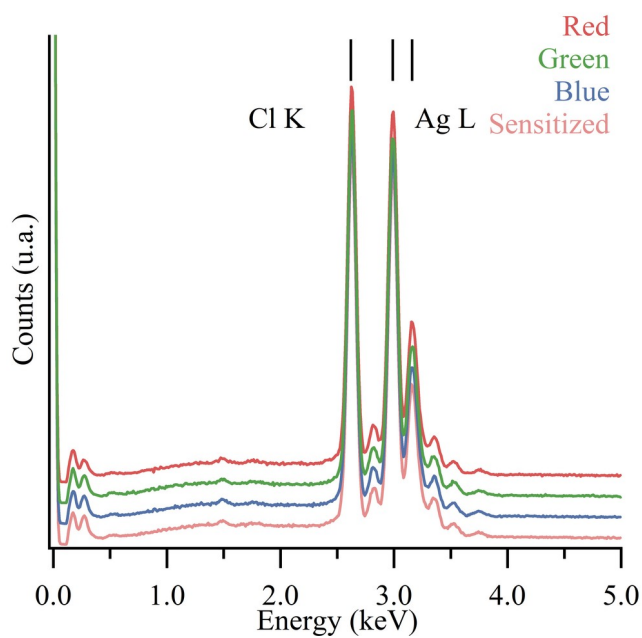

**Figure S-1.** EDX spectra of coloured model samples.

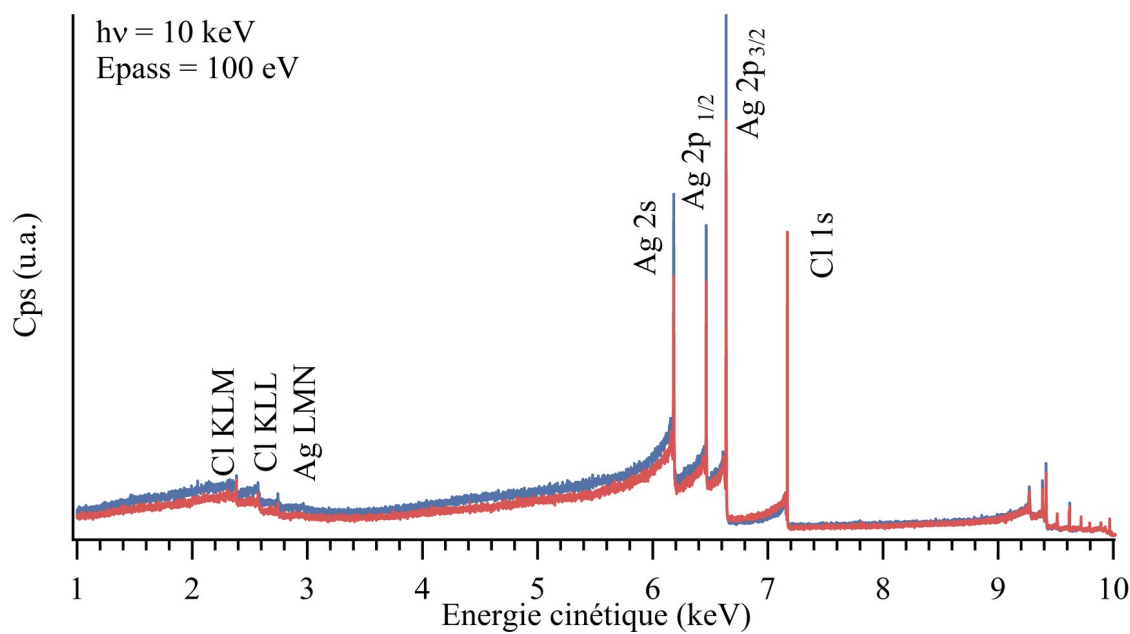

**Figure S-2.** HAXPES survey scan of coloured samples.

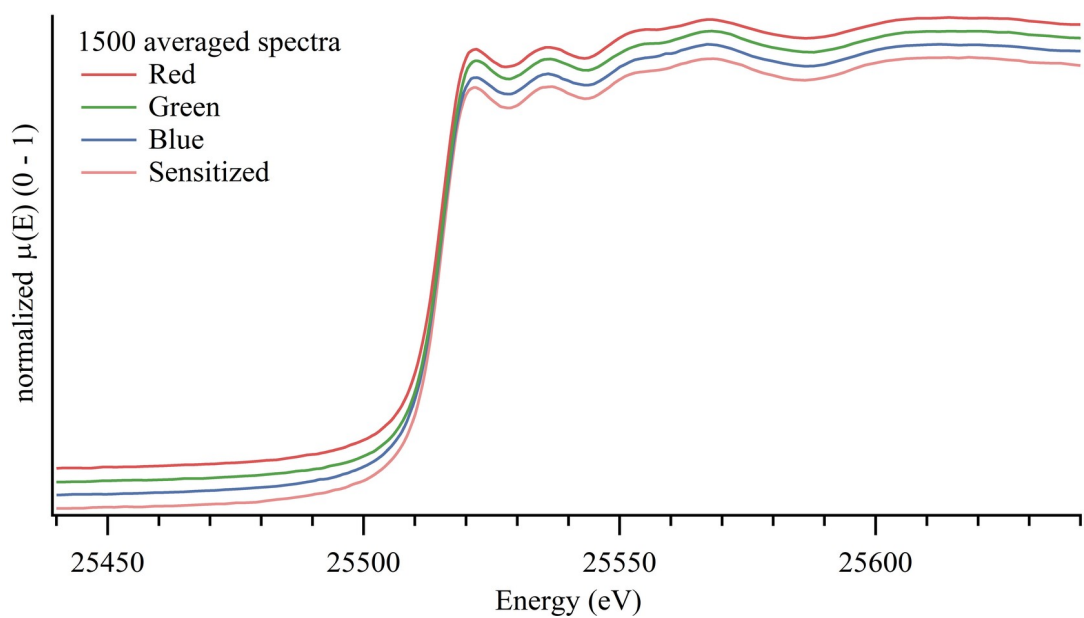

**Figure S-3.** XAS spectra of coloured samples.

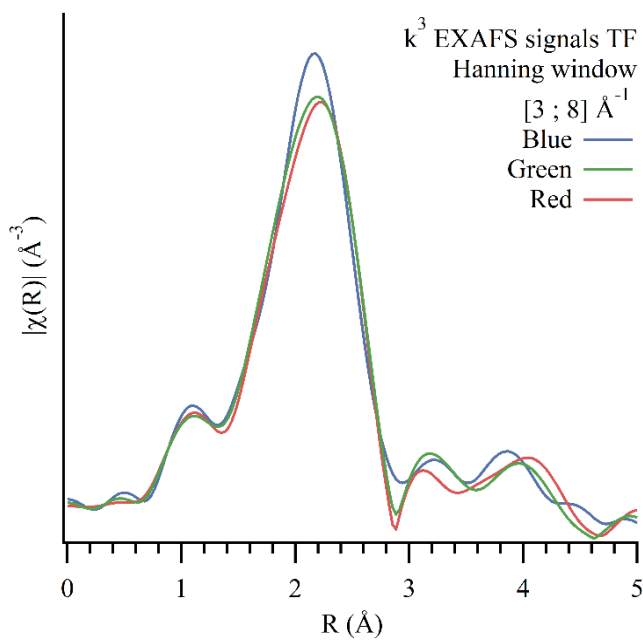

**Figure S-4.** FTs of the EXAFS signals of coloured model samples (1500 averaged spectra).

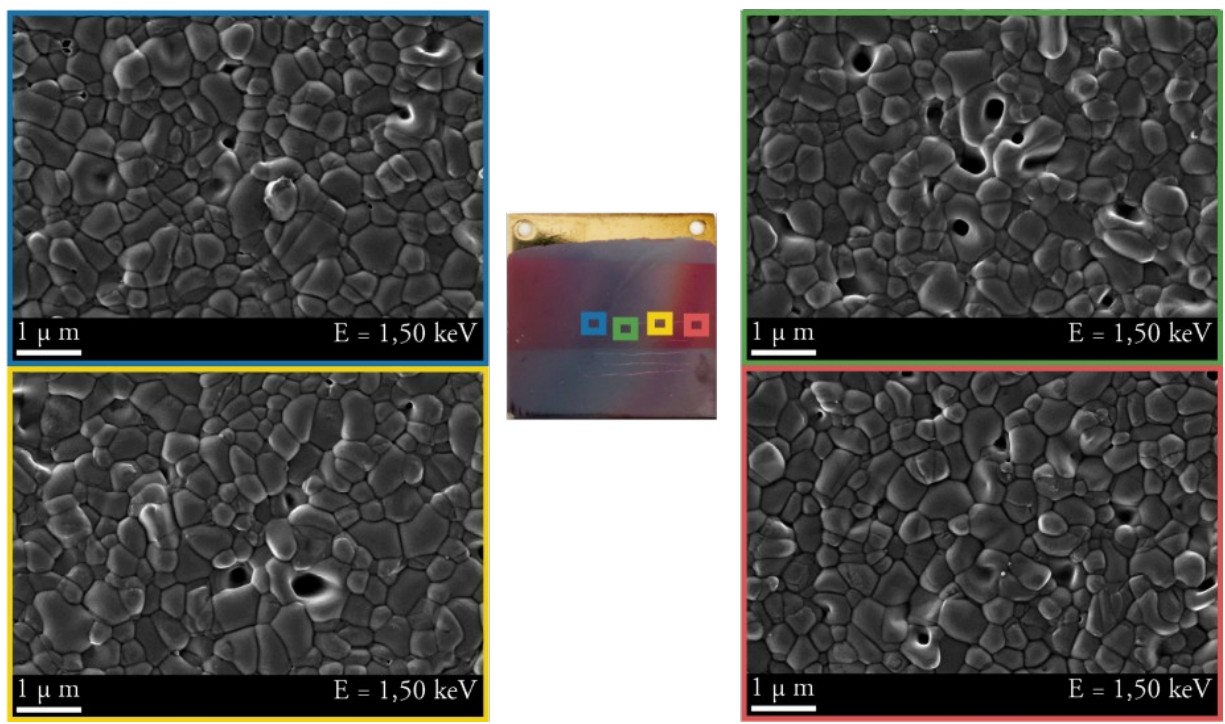

**Figure S-5.** Secondary electron images of the surface morphology of a sample reproducing a visible spectrum. Coloured areas were imaged and identified macroscopically by marks on the sample surface.

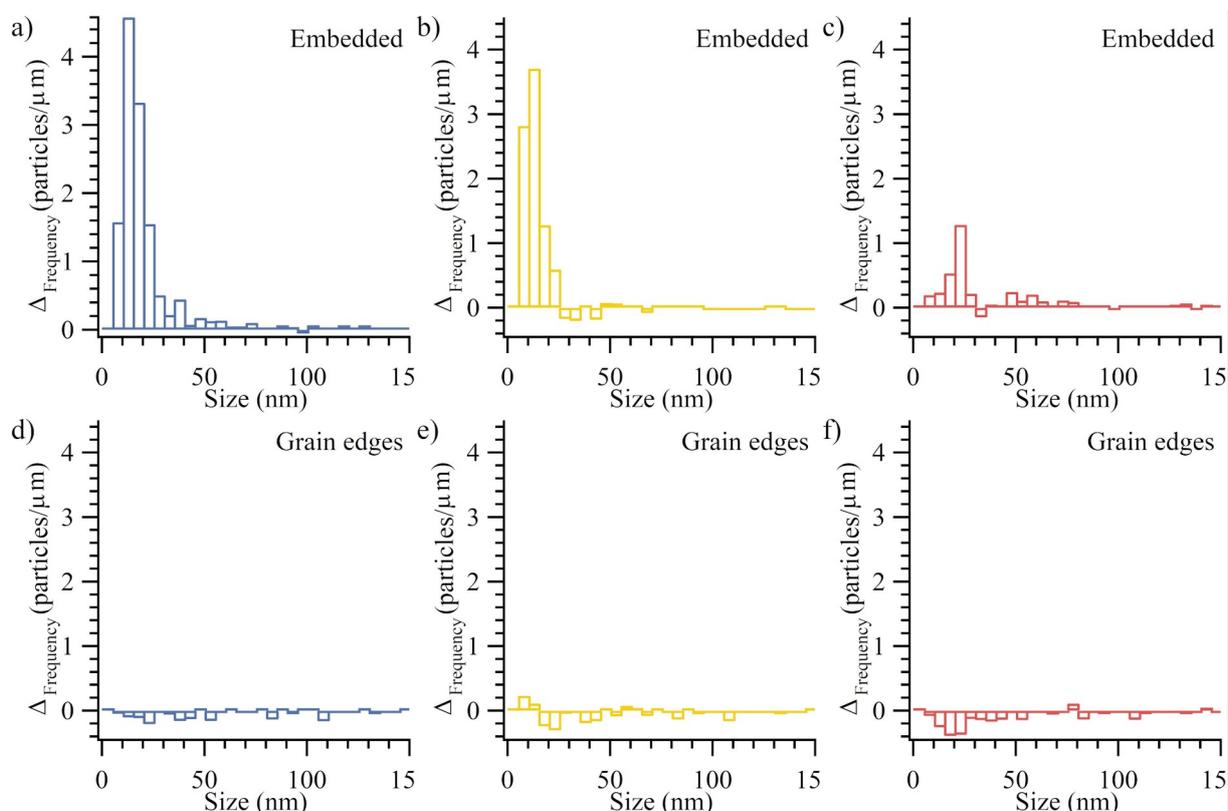

**Figure S-6.** Ag-NPs size distribution variations for several coloured samples, according to the localizations of the nanoparticles.

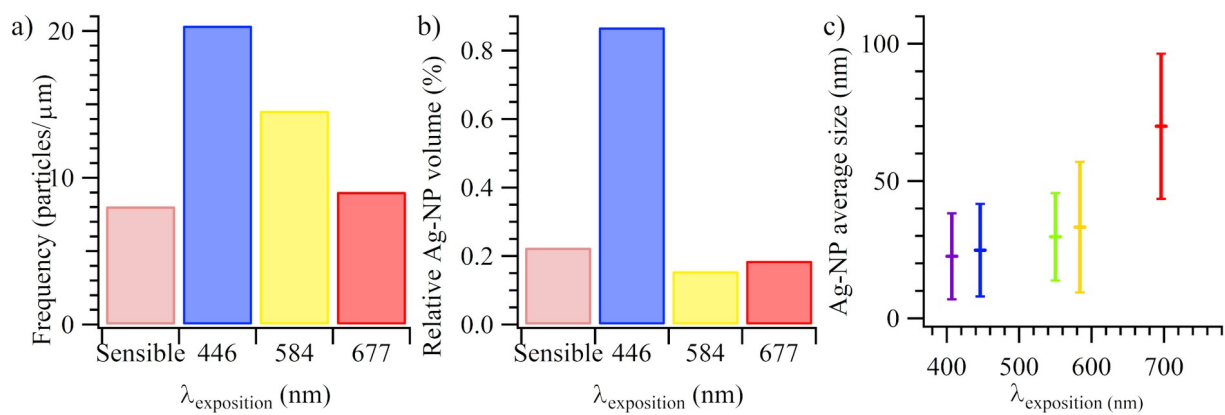

**Figure S-7.** Total number of silver nanoparticles per unit of layer length (a) and relative volume occupied by the silver nanoparticles (b) for sensitized and coloured samples. Average nanoparticles size versus the exposition wavelength used to colour the sample (c).

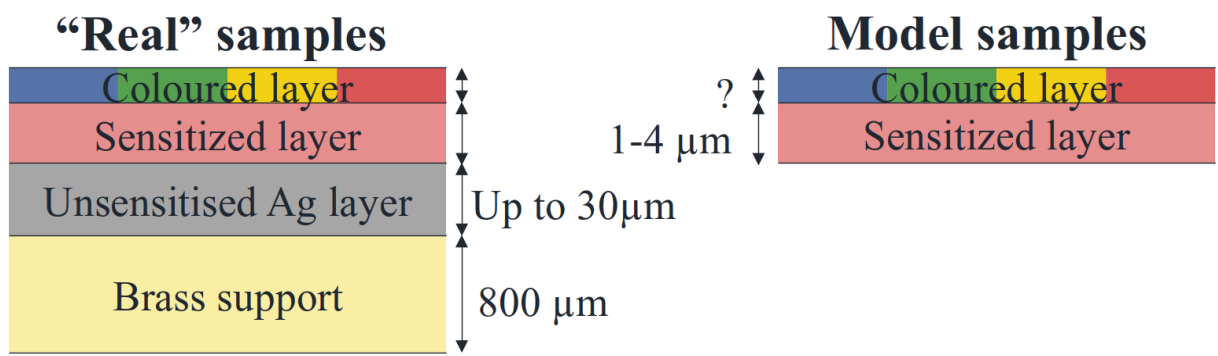

**Figure S-8.** Scheme of real and model samples.